**1. Title Page**

**Title of article:**

Human Lumbar Spine Injury Risk in Dynamic Combined Compression and Flexion Loading

**Authors:**

*Sophia K. Tushak[a], Bronislaw D. Gepner[a], Jason L. Forman[a], Jason J. Hallman[b], Bengt Pipkorn[c], Jason R. Kerrigan[a]*

**Institutions:**

*[a]Center for Applied Biomechanics, University of Virginia, Charlottesville, Virginia;*

*[b]Toyota Motor Engineering & Manufacturing North America, Inc., United States;*

*[c]Autoliv Research, Sweden*

**Abbreviated Title:**

Lumbar Spine Injury Risk in Compression and Flexion

**Correspondence Contact Information:**

<u>Author:</u> Sophia K. Tushak
<u>Affiliation</u>: Center for Applied Biomechanics, University of Virginia
<u>Address</u>: 4040 Lewis and Clark Drive, Charlottesville, VA, 22911
<u>Phone</u>: +1 (434) 297-8079
<u>Email</u>: skt5ay@virginia.edu

## 2. Abstract and Key Terms

Anticipating changes to vehicle interiors with future automated driving systems, the automobile industry recently has focused attention on crash response in relaxed postures with increased seatback recline. Prior research found that this posture may result in greater risk of lumbar spine injury in the event of a frontal crash. This study developed a lumbar spine injury risk function that estimated injury risk as a function of simultaneously applied compression force and flexion moment. Force and moment failure data from 40 compression-flexion tests were utilized in a Weibull survival model, including appropriate data censoring. A mechanics-based injury metric was formulated, where lumbar spine compression force and flexion moment were normalized to specimen geometry. Subject age was incorporated as a covariate to further improve model fit. A weighting factor was included to adjust the influence of force and moment, and parameter optimization yielded a value of 0.11. Thus, the normalized compression force component had a greater effect on injury risk than the normalized flexion moment component. Additionally, as force was nominally increased, less moment was required to produce injury for a given age and specimen geometry. The resulting injury risk function can be utilized to improve occupant safety in the field.

**Key terms**: combined loading, injury risk function, injury risk curve, motor vehicle crashes, frontal crashes, vertebral body fractures

**Word Count**:

Abstract – 199

Text – 5255, Introduction through References including all titles

**Number of Figures/Tables**: 4 figures, 1 table

### 3. Introduction

Human biomechanical and injury responses in novel driving postures have received increased attention with the ongoing development of automated driving systems (ADSs). One posture receiving increased attention is a reclined occupant. Although this posture may enable more comfort and convenience, prior computational human body modelling studies have suggested that the lumbar spine in this posture may experience higher combined axial compression and flexion than typically experienced by motor vehicle occupants in frontal crashes in the standard upright posture, resulting in the potential for lumbar spine injuries[2,4,8,9,11,12,14,22,28]. Lumbar spine fractures also have reportedly demonstrated an increasing frequency in crash field data[5,15,20,33,26], with the majority occurring in frontal crashes[20,26]. Additionally, lumbar spine injuries, including vertebral body fractures, have been reported in frontal full-body post-mortem human subject sled tests with upright[3,13,19,25,31] and reclined[23] postures. The compression and burst vertebral body fractures observed in frontal crashes have been attributed to combined compression and flexion and compression-only loading[1,10,24,29].

To assess the risk for injury at any given compression force and flexion moment, an injury risk function (IRF) that combines those structural response variables into a single input predictor may be utilized. In general, an IRF is a probabilistic model that relates an independent predictor variable, such as force, moment, or a function-based injury metric, to the probability of an injury occurring. Modifiers, such as age, sex, height, or weight, may be incorporated into the IRF as covariates. Age is often chosen as a covariate since it is generally correlated to injury tolerance and injury risk[21]. Survival analysis is a common statistical method for developing injury risk functions in injury biomechanics for its inclusion of data censoring. While the Weibull distribution has become increasingly popular for biomechanical injury risk analyses[16], it has been found that other distributions (logistic, log-logistic) can be appropriate, but Weibull is often favorable[17]. Ultimately, IRFs are applied to human surrogates, such as computational human body models, anthropomorphic test devices (i.e. crash test dummies), and post-mortem human subjects, to predict the risk of injury in motor vehicle crashes and to evaluate the benefits of potential countermeasures.

Previous studies investigating injury tolerance to combined loading of the cervical spine and tibia have proposed injury metrics which combined force and moment linearly after normalizing each by critical values[6]. While these metrics normalize measured forces and moments by one pre-determined critical force and moment to create unitless metrics, their approach has not been linked to the theory of stress superposition and its application to curved-beam analysis. Further, the effects of axial compression and flexion bending after normalizing by geometric parameters could be combined to obtain a metric that more closely represents the stress applied. Such a formulation may provide for more accurate injury prediction because it is supported by the fundamental mechanical principle of superposition and incorporates specimen-specific geometries with failure tolerance for greater detail.

The failure tolerance of the lumbar spine in combined compression and flexion was investigated in a previous experimental study[30]. However, the previous study did not attempt to use the failure tolerance data to examine injury prediction metrics or IRFs. Therefore, the objective of this study was to determine lumbar spine injury risk for combined compression and flexion loading by developing an injury risk metric and IRF that best fits previously published data.

**4. Materials and Methods**

The present study utilized previously reported data on experiments conducted on isolated ligamentous lumbar spines[30]. Briefly, forty 3-vertebrae lumbar spine segments (n=21 T12-L2, n=19 L3-L5; n=21 Male, n=19 Female) with average middle vertebrae cross-sectional area (CSA) of $973*10^6$ $m^2$ (range: $676*10^6$ $m^2$ to $1598*10^6$ $m^2$) and average donor age of 49 years old (range: 21 years old to 74 years old) were utilized. Vertebral body CSA was calculated from the superior endplate of each middle vertebra (L1 or L4) via CT scans. CSA was calculated as the area within a series of approximately 40 points along the outermost edge of the superior endplate. The superior and inferior vertebrae were secured, while the middle vertebrae and both intervertebral discs were unconstrained. A linear acceleration single intrusion cylinder was modified to deliver high-rate rotational motion. The boundary conditions applied by the test fixture permitted the application of pure flexion moment with superimposed axial compression force. Specimens were quasi-statically axially compressed to one of three nominal force levels (2200 N, 3300 N, or 4500 N), and then dynamically flexed at an average rate of 600 °/s to about 45°. In-depth analysis of acoustic sensor, strain gage, force, and moment data, as well as high-speed videos and pre- versus post-test computed tomography (CT) scans, led to the determination of failure timing, type, and censoring for each specimen. All failure loads were transformed (i.e. rotated and translated) from the inferior load cell coordinate system to the middle vertebrae coordinate system regardless of censoring.

For the purposes of the IRF development, injury was defined as endplate, wedge, compression, or burst fractures within the middle vertebra of the three-vertebrae segments. Average failure force was 3405 N (range: 1580 N to 5062 N), and average failure moment was 71 Nm (range: 0 Nm to 182 Nm). Failure data were characterized as uncensored, left-censored, or right-censored based on the nature of the injury. Uncensored specimens fractured in the middle vertebrae (L1, L4) at an exact, known time (18 specimens). Left-censored specimens fractured in the middle vertebrae during the quasi-static application of axial compression and only the upper limit of applied force was known (3 specimens). For the three left-censored specimens and three of the uncensored specimens that failed during the application of axial compression, failure was assumed to be in pure compression, and, thus, all other loads were assumed to be zero. Right-censored specimens fractured in the superior or inferior vertebral bodies or at the potting resin/bone interface (19 specimens). Middle vertebrae fractures were desired and fractures occurring in the superior and inferior vertebrae were likely influenced by the boundary effects, thus constituting the right-censoring classification.

A mechanics-based formulation was selected for the IRF input metric, in which the lumbar spine was modelled as a beam subjected to a combination of stresses from the axial compression ($F$) and flexion moment ($M$), producing a single input predictor variable, $L_{fx}$ (Eq. 1), to describe lumbar spine fracture criteria. The predictor equation is a linear combination of stress-like terms: normalized axial compression ($F/CSA$) and normalized flexion moment ($M/CSA^{3/2}$), resulting in units of N/m$^2$. $CSA^{3/2}$ was chosen to approximate the section modulus (with similar units of length$^3$), which was necessary since the neutral axis could not be confidently determined.

$$L_{fx}(t) = (1-\alpha)\frac{F(t)}{CSA} + \alpha\frac{M(t)}{CSA^{3/2}} \qquad \text{(Eq. 1)}$$

The predictor equation was optimized using a relative contribution factor, alpha (α), to produce the IRF with best predictive fit to the data. The purpose of α was to adjust the relative contributions of force and moment and account for errors between $CSA^{3/2}$ and section modulus. Due to the formulation of the equation, α was bounded between zero and one for simplicity and was increased in increments of 0.01. The optimized α yielding the best predictive fit was defined as the value which resulted in the highest log-likelihood. The IRF and α optimization were fit using survival analysis with a Weibull distribution, which incorporated differences in data censoring[16,17], with the main predictor variable, $L_{fx}$, and donor age as a covariate (R, v.4.0.3).

## 5. Results

The optimal α value was 0.11. The normalized force and moment components were originally of similar magnitude. However, the introduction and optimization of α reduced the normalized moment component by an order of magnitude (Figure 1), as evidenced by an alpha value of 0.11 equating to a multiplier of approximately 10%. Average $L_{fx}$ was $3.44*10^6$ N/m² (range: $1.77*10^6$ N/m² to $6.12*10^6$ N/m²). The probability of injury could be described as:

$$P(fracture|L_{fx}, Age) = 1 - e^{-\left[\frac{L_{fx}}{e^{\beta_0+\beta_1 Age}}\right]^{\lambda}},$$ 

(Eq. 2)

where $L_{fx}$ is the predictor variable in units of MPa (N/mm²), $\beta_0$ and $\beta_1$ are the coefficients of the intercept and age covariate, and λ is 1/(scale parameter) (Table 1). The resulting probability of fracture function for this data was:

$$P(fracture|L_{fx}, Age) = 1 - e^{-\left[\frac{L_{fx}}{e^{1.89043-0.00886*Age}}\right]^{\frac{1}{0.201}}}$$

(Eq. 3)

As age increased, less force and/or moment was required for injury to occur for a given fracture probability (Figure 2). The confidence intervals were wider at the 25-year-old level than at 45-year-old and 65-year-old levels. Force was solved over the range of $L_{fx}$ using Eq. 1 by setting CSA (average = 0.000973 m$^2$), age (25 years, 45 years, 65 years), moment (0 Nm, 100 Nm, 200 Nm), and α (0.11) to determine how force stimulus changed for different combinations of these variables. As force was nominally increased, less moment was required for injury to occur for a given fracture probability, CSA, and age (Figure 3). A similar process was performed to solve for moment at several levels of force (2200 N, 3300 N, 4500 N). Likewise, as moment was nominally increased, less force was required for injury to occur for a given fracture probability, CSA, and age (Figure 4).

## 6. Discussion

The goal of this study was to quantify lumbar spine injury risk and develop an IRF for combined compression-flexion loading using previously published failure tolerance data. Generally, more force and/or moment was required to injure younger specimens than older specimens. Likewise, when force was increased less moment was required for injury, and vice versa. The tails of the IRFs at the 25- and 45-year-old level did not extend substantially past the range of $L_{fx}$ data, and the IRF at the 65-year-old level was within the range of the experimental data. This instilled confidence that the IRF can predict the highest level of injury risk, and trust that the prediction increased as age increased. Similarly, the confidence intervals were wider at the 25-year-old level since 12 out of the 40 specimens were younger than 45 years old and four of the specimens were younger than 25 years old. Thus, injury risk was better captured for this dataset at the 45-year-old and 65-year-old levels due to the distribution of donor age in the data, while still representing injury risk for some younger occupants.

In its base form, the IRF had two degrees of freedom: α, which was fit using optimization, and the coefficient of $L_{fx}$. As a general guideline, having at least 10 injury and 10 non-injury data points for doubly censored data is necessary for a reliable fit[32]. Since this test series produced 40 data points, that would allow for 40 / (10 injury + 10 non-injury) = 2 degrees of freedom. However, seeing as almost half of the data is exact, or uncensored, there was potential to include one more degree of freedom without overfitting the model. Therefore, age was included as a covariate since it is generally related to bone properties (bone mineral density, cortical thickness, trabecular density), which were not included in the model formulation, and yielded a significant modifier of predicting injury risk ($p<0.01$). Age was also less likely to be correlated with CSA, which was already included in the model formulation. Thus, three model parameters were fit. Sex was also considered as a covariate, but the addition of sex yielded slightly worse fit to the data (quantified by AIC) and did not significantly affect injury risk prediction ($p>0.05$). Additionally, sex is generally more likely to be correlated to geometry, in which male CSA was substantially larger than female CSA in this dataset, and CSA was already included in the model. When choosing the CSA for the beam equation formulation, the superior endplate was preferred due to the ease and convenience of taking the measurement for experimenters and modellers. It should be noted that all 40 data points were treated as independent, and there was assumed not to be significant correlation between T12-L2 and L3-L5 specimens segmented from the same donor spine.

The primary benefit of including the weighting factor, α, was that it optimized the relative contributions of force and moment and adjusted for utilizing $CSA^{3/2}$ as opposed to section modulus when calculating flexion moment stress. Previous IRF predictor equations in which force and moment were in a linear combination and divided by critical values[6] assumed the relative contributions of force and moment. The weighting of one relative to the other were determined by the critical values. The weighting factor in this study evaluated the current dataset across a bounded range to determine the value that yielded the best injury risk prediction. For the flexion moment component, an arbitrary $CSA^{3/2}$, analogous to section modulus, was chosen since the location of the neutral axis at failure could not be calculated with confidence due to the composite nature of the spinal segment and its non-trivial stress state.

Furthermore, the relative contributions of force and moment, modified by α, have several implications. Originally, the inputs from the force ($F/CSA$) and moment ($M/CSA^{3/2}$) components were of the same order of magnitude. However, the optimized α reduced the inputs from the moment component by one order of magnitude (reduced to 11%), while the inputs from the force component remained relatively similar (reduced to 89%). The optimized α illustrated that 89% of the injury risk for any given stress state is due to normal stress and 11% of the injury risk is related to bending stress. Thus, the force component had a greater influence on predicting injury risk than moment component for this dataset.

The large reduction in the contribution from the flexion moment (low α value) may be due to the inaccuracy between $CSA^{3/2}$ and section modulus, and α may be the optimized correction for the incorrect relative magnitude between the two. Additionally, the relative strength of the lumbar spine anatomic components may have affected failure, and, thus, the optimized relative contributions of force and moment. The previous experiments[30] described injuries in which compression and compression-flexion failures, such as vertebral body endplate, wedge, compression, and burst fractures, occurred before any flexion-based failures, such as ligament injuries. Due to the nature of the injuries, compression played a major role in failure, with flexion playing a smaller, yet still important, role in failure. This relationship was portrayed in the α optimization since the flexion moment component contributed less to injury risk compared to the compression force component. Nonetheless, failure moment is necessary to include in the predictor variable formulation because the failure mechanism for the experimental tests was combined compression and flexion loading for all but six specimens (other six specimens failed in compression), as evidenced by the mechanical test fixture design and the large forces and moments at failure. Additionally, the statistical fit to the data improved when fitting the $L_{fx}$ versus fitting only the force component. Lastly, approximately half of the data were right-censored, describing specimens which failed in superior or inferior vertebrae or potting before the middle vertebrae. These data points provided valuable information about non-injury cases, as these specimens would have failed in the middle vertebrae at forces and/or moments higher than those reported.

In the experiments, the specimens recorded non-negligible off-axis loads at the time of failure, particularly AP shear forces (0 to 2154 N) and lateral bending moments (0 to 41 Nm). The AP shear forces in the middle vertebrae were attributed to force transformation from the load cell to the middle vertebrae. The lateral bending moments were attributed to specimen medial-lateral asymmetry or scoliosis. As an exercise, component and resultant forces and moments were implemented into the predictor function (Eq. 1) to evaluate how the addition of off-axis loads affected injury prediction and model fit. In addition to the previously described formulation using component loads (compression force and flexion moment), the following models were assessed: resultant sagittal force (compression and AP shear) and flexion moment, compression force and resultant moment (flexion and lateral bending), and resultant sagittal force and resultant moment. Injury prediction and logliklihood did not largely vary by component versus resultant forces and moments. For the same injury risk level, $L_{fx}$ values varied by less than 7% of the baseline formulation (compression force and flexion moment), the differences of which are negligible in comparison to the larger confidence intervals. Thus, it is arguable that any type of input forces and moments may be feasible. However, compression force and flexion moment are the most appropriate when determining stresses in a beam due to the applied loading from these experiments and the resulting clinically relevant fractures, and are therefore suggested to be used over resultant loads. Similarly, logistic and loglogistic distributions were also considered, but the Weibull distribution yielded better fit, as measured by Akaike information criterion (AIC).

Several assumptions in this study were employed. First, modelling the lumbar spine as a beam was a simplification. In reality, the lumbar spine is a complex composition of stacked vertebrae with intervertebral discs providing shock absorption and ligaments and tendons providing constraint of movement. However, the weighting factor dictated that the relative contribution from the moment component was substantially less than that from the force component. A similar finding was described previously[18], who implemented an approach similar to the cervical spine and tibia to predict lumbar spine injury risk using a linear combination of force and moment divided by critical values. They found that the optimized critical value associated with moment (1155 Nm) was one to two degrees of magnitude larger than reported flexion moments, resulting in the minimization of its contribution in predicting injury risk. For example, the 50% injury risk corresponds to $L_{fx}$ of 3.5 for the 65 year-old level and average CSA. For comparison, in Ortiz-Paparoni et al.[18] 50% injury risk corresponds to a $\kappa$ of 1.0, and average donor age was 66 years old. After setting moment equal to the average failure moment from combined compression-flexion loading[30] in both equations (71 Nm), of which was within the range of flexion moments for both studies, forces of 5456 N and 3512 N were required to reach 50% injury risk probability using $\kappa$ and $L_{fx}$ equations respectively. Thus, $L_{fx}$ depends on a higher contribution from flexion moment in predicting fractures than $\kappa$, given the same flexion moment and comparable donor age. This could be explained by the loading in each study. Ortiz-Paparoni et al. [18] prescribed axial compression, while Tushak et al.[30] prescribed combined compression and flexion. Therefore, when considering injury risk of the lumbar spine in motor vehicle crashes, it is critical to use injury risk prediction tools which represent the desired loading scenario (i.e. combined compression and flexion) in order to avoid underestimation of injury risk for a given compression force and moment. One thing to note is that there are differences in censoring and size and sex of spinal segments utilized among both experimental studies, which could affect the resulting failure tolerance.

Nonetheless, it is advised that the IRF described in this study only be used in similar loading scenarios to the experiments in which it was fit. Thus, the IRF is not applicable to predicting injury risk in pure moment (with the absence of compression force). In the experiments, the specimens were quasi-statically compressed to one of three nominal force values, and then dynamically flexed beyond failure. Considering all of the tests, axial compression forces and flexion moments at fracture were always higher than approximately 1500 N and 0 Nm and less than approximately 5100 N and 180 Nm. While very low or zero axial compression forces can be implemented in the injury risk function, loading applied to the vertebrae, and the resulting injury mechanism, with pure flexion loading is likely different from injuries observed in the study. Therefore, the authors believe that the IRF presented in this study is limited to axial compression forces greater than approximately 1500 N and less than approximately 5100 N (Figure 4) to ensure that injury predictions are representative of the data used to develop the IRF. Likewise, the IRF may not be appropriate for flexion moments well above the upper bound of moment from the experiments (181 Nm) (Figure 3). It may seem like the lumbar spine can withstand upwards of 1000 Nm with 2200 N of compression force (Figure 3), but there is no data to suggest that this is true. The IRF's realm of applicability remains within or close to the bounds of the experimental data used to generate the IRF.

Alternatively, the IRF can be used in pure compression cases (with the absence of flexion moment) since the experiments included six specimens that failed in pure compression before flexion moment could be applied. Likewise, injury risk prediction using only compression force in the IRF from this study were similar to those predicted using an existing IRF with compression force as the input predictor[27]. Both studies had similar average donor age of 49 years old with a wide range of ages, included both males and females, and produced similar fracture types. The resulting IRF predicted that the force required for 50% injury risk was 4500 N, which was approximately the same force of 4473 N required to reach 50% injury risk for the IRF at the 45-year-old, 0

Nm level from this study when using the average CSA (Figure 4). Similarly, when only considering force as an input, another study[18] predicted 50% risk of injury for forces of approximately 5000 N. On the other hand, a third study[34] predicted 50% injury risk at forces of around 6000 N, which could be higher due to testing only male specimens, of which had larger vertebral body CSA than the specimens included in this study.

Another assumption was that for the right-censored data, it was assumed that the middle vertebrae would have fractured if higher forces and moments were successfully applied. Instead, the specimens failed in the potted vertebrae at lower forces and moments that were ultimately transformed from the inferior load cell to the middle vertebrae. It can be observed that $L_{fx}$ is within the range of approximately 1.5 to 8 N/m$^2$ regardless of censoring. The overlap of uncensored, left-censored, and right-censored data is likely due to the wide variation in donor age, sex, and anthropometry and the presumable wide variation in bone and ligament material properties. The aforementioned assumption that higher loads would be required to injure the middle vertebrae is valid on a specimen-by-specimen basis, with the necessity to take these varying characteristics into account.

The proposed IRF for compression and flexion loading to the lumbar spine may be useful for future research in lumbar spine injuries in motor vehicle crashes in which occupants typically experience initial axial compression followed by flexion during forward torso pitch in frontal crashes. Future ADS-equipped vehicles may offer occupants the opportunity to assume a more reclined posture when freed from the driving task. This reclined posture was previously reported to exhibit a larger magnitude of compression force within the lumbar spine prior to flexion during a crash compared to more upright postures. In addition, prior field studies and full-body post-mortem human subject sled tests have reported lumbar spine injuries in frontal crashes. Therefore, further research with this IRF may enhance safety in both novel postures suggested by ADS development trends and traditional driving postures. Although the IRF was developed using failure data, it can be applied in circumstances in which exact failure loads may not be known or cannot be determined. Specifically, force and moment may be measured over the duration of the experiment or simulation, input to calculate $L_{fx}(t)$, and then maximum $L_{fx}$ may be chosen to predict maximum probability of lumbar spine injury.

The IRF developed in this study is the first IRF, to the authors' knowledge, developed for combined compression and flexion loading of the lumbar spine similar to that observed in motor vehicle crashes. The IRF predicts lumbar spine vertebral body fracture risk as a function of a predictor variable that superposes normalized axial compression and normalized flexion moment. The relative contributions of force and moment were optimized such that the resulting IRF provides a best-fit model to the data. The proposed IRF may be used to predict lumbar spine injury risk in frontal motor vehicle crashes with human surrogates, with some possible necessary adaptations depending on the surrogate (e.g. anthropomorphic test devices, human body models). Ultimately, the IRF has the potential to direct research toward improving vehicle safety in future traditional and ADS vehicles.

**Conflicts of Interest**

The authors do not have any conflicts of interest to disclose.

# 7. Acknowledgments

The Toyota Collaborative Safety Research Center and Autoliv provided funding and technical support for this study. It should be noted that the views or opinions expressed here are those of the authors and not necessarily aligned with the views and opinions of the sponsoring organizations.

**Figures and Tables**

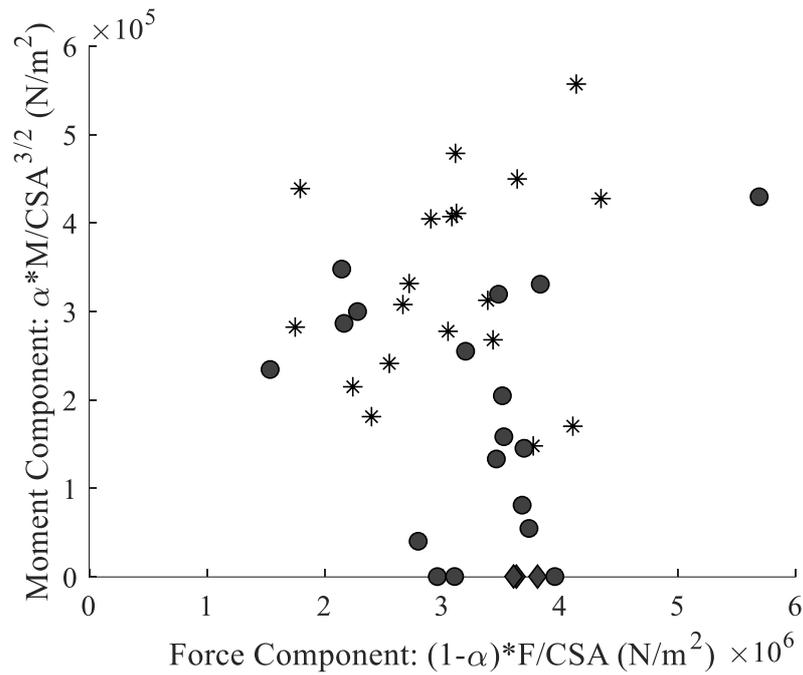

(Appears on page 8) Figure 1. Distribution of the force component (($1-\alpha$)*F/CSA) and moment component ($\alpha$*M/CSA$^{3/2}$) by censoring: uncensored (circle), left-censored (diamond), and right-censored (asterisk).

(Appears on page 9) Table 1. Parameter outputs from the survival analysis.

| Constant | $\beta_0$ (Intercept) | $\beta_1$ (Age) | Scale |
|---|---|---|---|
| Value | 1.89043 | -0.00886 | 0.201 |

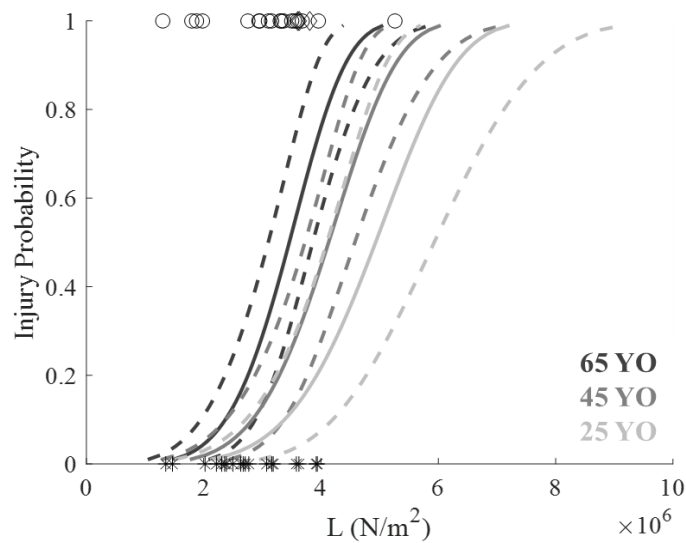

(Appears on page 9) Figure 2. Injury probability as a function of the main predictor variable, $L_{fx}$, using average CSA and separated by three age levels. Mean with 95% confidence intervals. Data points by censoring: uncensored (circle), left-censored (diamond), and right-censored (asterisk). Injury Probability = 1 corresponds to injury and Injury Probability = 0 corresponds to non-injury.

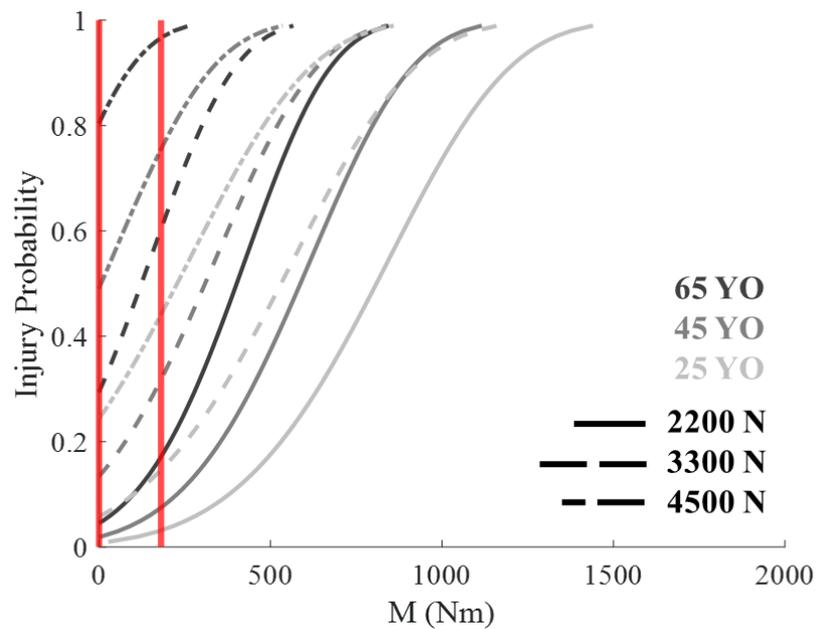

(Appears on page 9) Figure 3. Injury probability as a function of moment, $M$, using average CSA and separated by three age levels and three nominal force levels. Mean with 95% confidence intervals. Vertical lines at minimum and maximum moments measured from the experiments (0, 181 Nm).

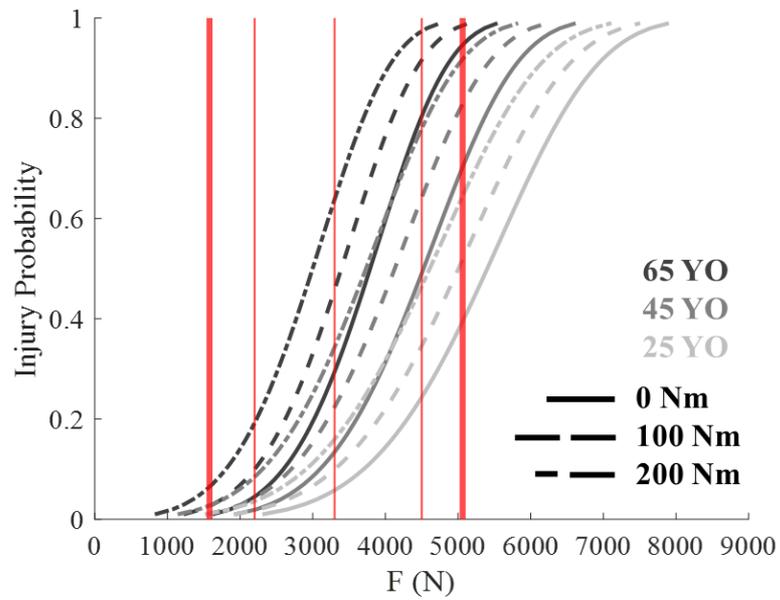

(Appears on page 9) Figure 4. Injury probability as a function of force, *F*, using average CSA and separated by three age levels and three moment levels. Mean with 95% confidence intervals. Vertical lines at minimum and maximum forces measured from the experiments (1580, 5062 N) and at the three nominal force levels (2200, 3300, 4500 N).